\documentclass[preprint,3p,times,numbers,sort&compress]{elsarticle}   
\usepackage{mathtools}
\usepackage{bm}	
\usepackage{caption}
\usepackage{subcaption}
\usepackage{siunitx}
\usepackage{booktabs}
\usepackage[english]{cleveref}
\usepackage{lineno}
\usepackage{float}
\usepackage{graphicx}
\usepackage{verbatim}
\usepackage{bibentry}

\def\tsc#1{\csdef{#1}{\textsc{\lowercase{#1}}\xspace}}
\tsc{WGM}
\tsc{QE}
\tsc{EP}
\tsc{PMS}
\tsc{BEC}
\tsc{DE}
\usepackage{color}  
\newcommand{\blue}{\color{black}} 

\newcommand{\black}{\color{black}}

\journal{Journal of Wind Engineering \& Industrial Aerodynamics} 

\begin{document}

\begin{frontmatter}

\title{A short note on turbulence characteristics in wind-turbine wakes}

\author{Navid Zehtabiyan-Rezaie}
\author{Mahdi Abkar\corref{cor1}}
\cortext[cor1]{Corresponding author}
\ead {abkar@mpe.au.dk}

\address{Department of Mechanical and Production Engineering, Aarhus University, 8200 Aarhus N, Denmark}

\begin{abstract}
Analytical wake models need formulations to mimic the impact of wind turbines on turbulence level in the wake region. Several correlations can be found in the literature for this purpose, one of which is the formula proposed in A. Crespo, J. Hern\'andez, Turbulence characteristics in wind-turbine wakes, Journal of Wind Engineering and Industrial Aerodynamics 61 (1) (1996) 71 – 85, which relates the added turbulence to the induction factor of the turbine, ambient turbulence intensity, and normalized distance from the rotor through an equation with one coefficient and three exponents for the effective parameters. Misuse of this formula with an incorrect exponent for the ambient turbulence intensity is propagating in the literature. In this note, we implement the original and the incorrect formulation of turbine-induced added turbulence in a Gaussian wake model to quantify its impact by studying the Horns Rev 1 wind farm. The results reveal that the turbulence intensity and the normalized power of the waked turbines predicted by the wake model with the correct and the incorrect implementation of turbine-induced added turbulence correlation have a difference equal to 1.94\% and 3.53\%, respectively, for an ambient turbulence intensity of 7.7\%. For an ambient turbulence intensity of 4\%, these discrepancies grow to 2.7\% and 4.95\%.
\end{abstract}


\end{frontmatter}

\section{Problem statement} \label{Sec:Introduction}
Wake models need to include the impact of wind turbines on turbulence level in the wake region. The turbulence intensity in the wake region can be defined as $I_{\text{w}} = \sqrt{I_0^2+I_+^2}$, where $I_0$ denotes the ambient turbulence intensity, and $I_+$ corresponds to the added turbulence intensity due to the presence of wind turbines. \blue Several studies have been conducted to examine various correlations of turbine-induced added turbulence, yielding valuable insights and recommendations, e.g., Refs. \cite{Crespo1996,Frandsen2007,Xie2015, Ishihara2018}, among others. \black 
One of the most widely used formulations to parameterize $I_+$ was suggested in the paper of Crespo and Hern\'andez \cite{Crespo1996} in 1996. They proposed calculating the maximum added turbulence intensity in the far-wake region with $0.07<I_0<0.14$ through

\begin{align}
     I_{+,\text{org.}} = 0.73 a^{0.8325} I_0^{-0.0325} \left({x}/d_0\right)^{-0.32},
    \label{eq:Crespo}
\end{align}

\noindent where $a$, $x$, and $d_0$ denote the axial induction factor of the turbine, the streamwise location, and the rotor diameter, respectively. The scanned paper of Crespo and Hern\'andez \cite{Crespo1996} is available, but the minus sign beside the exponent of $I_0$ is not clearly visible \blue (see Eq. (21) in the original paper)\black. This issue has found its way to many works, e.g. Refs. \cite{Niayifar2016, Gao2016, Bastankhah2021,Lopes2022,Li2022,FLORIS,pywake2023}, among others, which have a correlation implemented in their structure as

\begin{align}
     I_{+}= 0.73 a^{0.8325} I_0^{0.0325} \left({x}/d_0\right)^{-0.32}.
    \label{eq:CrespoIncorrect}
\end{align}

The goal of this short note is to clarify some confusion that seems to propagate in the literature about the misuse of the added turbulence correlation proposed by Crespo and Hern\'andez \cite{Crespo1996}. 
In connection with wind-farm flow modeling, the wake recovery rate in more recent and widely used analytical wake models \cite{Niayifar2016,Bastankhah2021} depends on the local turbulence intensity estimated by the above-mentioned formula.
Therefore, quantification of uncertainties associated with the implementation form of turbine-induced added turbulence is of high importance for the wind-energy community. 
Note that there exist few studies in the literature, e.g. \blue Refs. \cite{Vermeer2003,Xie2015, zehtab2022PRX}, among others\black, that utilized the correct correlation of Crespo and Hern\'andez \cite{Crespo1996} for estimation of the added turbulence intensity. 

\section{Methodology and discussion} \label{Sec:Analysis}
We utilize the analytical-empirical Gaussian wake model proposed by Niayifar and Port\'e-Agel \cite{Niayifar2016} as the main framework with the original (Eq.~(\ref{eq:Crespo})) and the incorrect implementation (Eq.~(\ref{eq:CrespoIncorrect})) for the turbine-induced added turbulence. The analytical-empirical Gaussian wake model is described in Ref. \cite{Niayifar2016}, and for the sake of brevity, we do not provide the detail on that.
We apply the wake model to the operational Horns Rev 1 (HR1) wind farm to evaluate the quantities of interest, e.g. the turbulence intensity within the farm and the normalized power, in the full-wake condition. The HR1 wind farm with a rhomboid-shaped layout holds eighty Vestas V-80 2MW turbines with hub height ($z_\text{h}$) and $d_0$ of 70 m and 80 m, respectively. For the full-wake condition, i.e. a wind direction of $270^\circ$, the spacing between consecutive turbines is $7d_0$ \cite{Wu2015}.
Even though this study does not aim to compare the accuracy of the wake model with different added-turbulence formulations, we present the predictions from large-eddy simulation (LES) \cite{Wu2015} for a hub-height velocity and ambient turbulence intensity of 8 m/s and 7.7\% in the following figures.

Figure~\ref{fig:TI_vs_Row} depicts the turbulence intensity obtained from LES \cite{Wu2015}, the correlation of Crespo and Hern\'andez \cite{Crespo1996} and its incorrect implementation, i.e. as Eq.~(\ref{eq:CrespoIncorrect}). The layout of the HR1 wind farm is also shown in this figure on the right. 
\blue The correlation of Crespo and Hern\'andez \cite{Crespo1996} gives the maximum added turbulence which appears near the top-tip and the two side-tip regions downwind of the turbine associated with the strong shear at those locations \cite{abkar2015influence,Abkar2016b,eidi2021model}.
In contrast, LES \cite{Wu2015} provides turbulence intensity averaged across the rotor, immediately upstream of each turbine. Consequently, the turbulence intensity calculated using Eq.~(\ref{eq:Crespo}) tends to be higher than LES data \cite{Wu2015}.
Turning to Eq.~(\ref{eq:CrespoIncorrect}), due to the missing minus sign beside the exponent of the ambient turbulence intensity, this equation yields smaller values compared to the correlation of Crespo and Hern\'andez \cite{Crespo1996}. 
\black
In a case with an ambient turbulence intensity of 7.7\%, we can see an absolute difference of 1.94\% between the turbulence intensity of the waked turbines predicted by the correct and the incorrect added-turbulence formulation. For a lower ambient turbulence intensity of 4\%, the difference grows to 2.7\%. 

\begin{figure}[ht]
	\centering
	\includegraphics[width=1\textwidth]{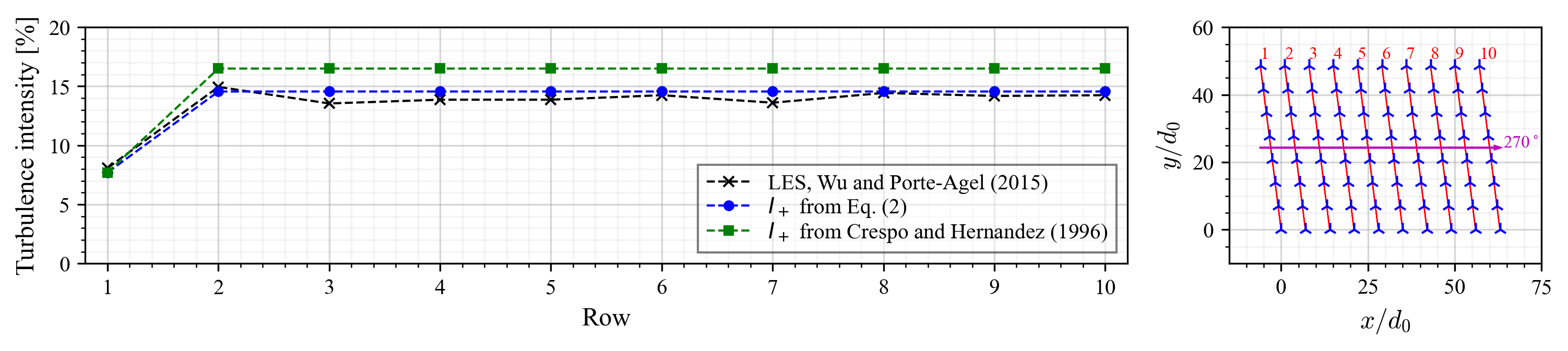}
	\caption{Left: The turbulence intensity averaged across the rotor upstream of each turbine row, derived from LES \cite{Wu2015}, compared to the values obtained using both the correct and incorrect implementations of the correlation introduced by Crespo and Hern\'andez \cite{Crespo1996}. The inlet hub-height velocity and ambient turbulence intensity are 8 m/s and 7.7\%, respectively. Right: The turbines marked with the red line are used to calculate the average turbulence intensity of each row.}
	\label{fig:TI_vs_Row}
\end{figure}

\begin{figure}[ht]
	\centering
	\includegraphics[width=1\textwidth]{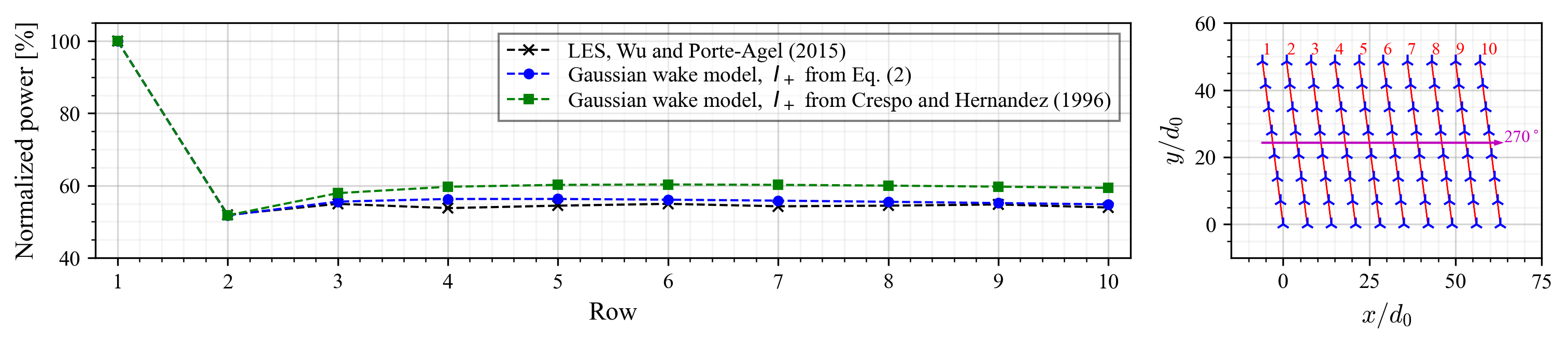}
	\caption{Left: The normalized mean power as a function of turbine row for a hub-height velocity and ambient turbulence intensity of 8 m/s and 7.7\%, respectively. Right: The turbines marked with the red line are used to calculate the average power of each row, and the average value of the first row is used as the reference to calculate the normalized power through $P_\text{row}/P_1 \times 100$. }
	\label{fig:NP_vs_Row}
\end{figure}

Figure~\ref{fig:NP_vs_Row} presents the normalized power for different rows in an incoming wind direction of $270^\circ$. As expected, the wake model \cite{Niayifar2016} using the original formulation for the added turbulence \cite{Crespo1996} predicts higher values for the normalized power due to a larger turbulence intensity in the wake and, consequently, a faster wake recovery. By using Eq.~(\ref{eq:CrespoIncorrect}) in the wake model \cite{Niayifar2016}, the average efficiency of the seventy-two waked turbines has an absolute difference of 3.53\% compared to the value obtained from the model with the original correlation. The difference is equal to 4.95\% for an ambient turbulence intensity of 4\%. 

\blue 
\section{Conclusions} \label{Sec:Conclusions}
The purpose of this short note was to address the confusion prevalent in the existing literature regarding the incorrect utilization of the added-turbulence correlation introduced by Crespo and Hern\'andez \cite{Crespo1996}. To this end, we employed the analytical-empirical Gaussian wake model proposed by Niayifar and Port\'e-Agel \cite{Niayifar2016} as the primary framework. Then, we proceeded to analyze both the original implementation (Eq.~(\ref{eq:Crespo})) and the incorrect implementation (Eq.~(\ref{eq:CrespoIncorrect})) of the turbine-induced added turbulence within this model. We applied the wake model to the HR1 wind farm under full-wake conditions and assessed two quantities of interest, i.e., the turbulence intensity in the wake region and the normalized power of the waked turbines. The results showed that employing the original formulation
leads to a higher normalized power compared to the value predicted by the wake model coupled with Eq.~(\ref{eq:CrespoIncorrect}). This was attributed to the increased turbulence intensity within the wake region, resulting in a faster wake recovery.
\black

\section*{CRediT authorship contribution statement}
\textbf{Navid Zehtabiyan-Rezaie}: Data curation, Formal analysis, Software, Writing – original draft. \textbf{Mahdi Abkar}: Formal analysis, Project administration, Resources, Supervision, Writing – review \& editing.

\section*{Declaration of competing interest}
The authors declare that they have no known competing financial interests or personal relationships that could have appeared to influence the work reported in this paper.

\section*{Data availability}
Data will be made available on request.

\section*{Acknowledgments}
The authors acknowledge the financial support from the Independent Research Fund Denmark (DFF) under the Grant No. 0217-00038B.

\bibliographystyle{elsarticle-num} 

\end{document}